# Ultra-low-power Monostatic Backscatter Platform with Phase-Aware Channel Estimation and System-Level Validation

Hanyeol Ryu and Sangkil Kim, *Senior Member, IEEE*

*Abstract*— This paper presents a novel channel-estimation (CE) method that mitigates residual phase drifts in backscatter links and a full hardware and signal-processing pipeline for a single-antenna monostatic system. The platform comprises a semi-passive tag, a software-defined radio (SDR) reader, and a 2×1 planar Yagi–Uda array (7 dBi with higher than 30 dB isolation) operating at 2.4 ~ 2.5 GHz. The developed backscatter fading model accounts for round-trip propagation and temporal correlation, and employs an analytically derived resource-optimal pilot allocation strategy. At the receiver, optimized least square (LS) and linear minimum mean square error (LMMSE) CE with pilot-aided carrier frequency offset (CFO) compensation feed a zero-forcing (ZF) equalizer to suppress ISI. The prototype delivers 500 kbps at 1 m with power of 158 µW (SDR baseband) and 10 µW (RF switch), yielding 320 pJ/bit. OOK and BPSK modulations achieve measured EVMs of 2.97 % and 4.02 %, respectively. Performance is validated by BER measurements and successful reconstruction of a full-color image in an over-the-air test. The results demonstrate an ultra-low-power, multimedia-capable backscatter IoT link and provide practical hardware–software co-design guidance for scalable deployments.

## I. INTRODUCTION

THE rapid growth and pervasive adoption of the Internet of Things (IoT) impose unprecedented demands on wireless communication technologies. Meeting these demands calls for solutions that are scalable, energy-efficient, and economically sustainable. Conventional wireless systems depend on active RF transceivers that consume significant power, are dependent on batteries, and incur high operational and maintenance costs. As these constraints hinder scaling IoT, the pursuit of ultra-low-power wireless solutions, particularly backscatter communication, has emerged as a central research direction for near-zero-power connectivity. By passively modulating incident or dedicated RF signals, backscatter devices drastically reduce energy consumption and enable battery-free or near-battery-free operation. This transformative capability establishes backscatter as a foundational technology for the expanding IoT and emerging 6G era, both of which are projected to connect devices at unprecedented scale. Such exponential growth further underscores the need for scalable, ultra-low-power communication methods that can meet the connectivity and maintenance requirements of next-generation IoT ecosystems.

### A. Motivation

Most prior backscatter systems have been confined to ultra-low-throughput applications such as radio-frequency identification (RFID) and basic sensing. The emerging demand for multimedia communication—ranging from high-resolution video surveillance and real-time biomedical telemetry to intelligent transportation and augmented-reality services—introduces far more stringent requirements in the throughput, latency, and reliability [1], [2], [3]. In particular, smart city infrastructures exemplify the growing demand for multimedia-oriented backscatter connectivity: such capabilities could enhance public safety, optimize traffic flow, and enable personalized healthcare, thereby transforming urban environments.

Meeting these requirements is impeded by several fundamental challenges. Backscatter channels are inherently more susceptible to deep fades, which undermines reliability, residual carrier phase distortions compromise channel estimation (CE) accuracy, and the absence of dedicated energy sources constrains sustainable operation. Furthermore, large-scale deployment would necessitate robust interference suppression, efficient pilot allocation, and infrastructure capable of integrating backscatter devices with edge and cloud intelligence. These gaps highlight the need for new system designs that move beyond basic identification and enable robust multimedia transmission while operating within strict power constraints.

### B. Related Work

Recently reported research efforts that have advanced and expanded the capabilities of backscatter communication are discussed in this subsection. These studies span multiple directions, including throughput enhancement, clarification of backscatter channel properties, and interference suppression. Collectively, these works show how backscatter has advanced

This work was supported by the National Research Foundation of Korea (NRF) Grant funded by the Korea Government (MSIT) under Grant RS-2023-00237172 (*Corresponding author: Sangkil Kim*).

Hanyeol Ryu and Sangkil Kim are with the Department of Electronics Engineering, Pusan National University, Busan 46241, Republic of Korea (e-mail: ksangkil3@pusan.ac.kr).

beyond its initial role in identification and sensing, while highlighting the ongoing challenges that demand further study.

*1) Throughput Enhancement and Higher-order Modulation:* Early backscatter systems showed simple amplitude shift keying (ASK) or phase shift keying (PSK) modulation. Building on these schemes, higher-order schemes such as 16-quadrature amplitude modulation (QAM) have been introduced, improving both spectral efficiency and data rates [4], [5]. Recent studies have extended this line of work to mmWave gigabit backscatter front-ends [6], [7]. In addition, optimized high-order modulation strategies based on transfer learning have been proposed to enhance dual-channel multiple-input multiple-output (MIMO) systems [8]. These developments position backscatter as a technology capable of evolving beyond identification and sensing into high-throughput, multimedia communication.

*2) Backscatter Channel Modeling and Estimation:* Foundational measurement studies revealed that small-scale fading in backscatter links behaves as the product of two independent channels, effectively doubling the path loss exponent [9]. Subsequent models have characterized spatial fading in both monostatic and bistatic scenarios [10], while retrodirective architectures were shown to alleviate the double-fading problem, restoring single-hop channel characteristics [11]. On the estimation side, approaches such as least squares and discrete Fourier transform (DFT)-based techniques for large antenna readers [12], reciprocity-driven full-duplex estimation [13], and low-complexity schemes employing reconfigurable surfaces [14] have been developed to improve robustness. These methods form the foundation for acquiring reliable channel state information (CSI) in both ambient and dedicated backscatter systems.

*3) Interference Cancellation and Reliability:* Another important research direction addresses interference, which is especially critical in monostatic and bistatic architectures where strong direct-link or self-interference can overshadow the weak backscattered signal. Hardware solutions include carrier suppression loops in UHF RFID readers [15], tunable isolation structures for co-polarized antennas [16], and enhanced metasurface monostatic designs [17]. On the algorithmic side, interference cancellation techniques have been developed for bistatic networks [18], while ambient backscatter using digital video broadcasting (DVB) signals has demonstrated practical interference rejection [19]. More recent works introduced likelihood-based detectors for distributed MIMO systems [20] and optimized methods to guarantee reliability under channel uncertainty and fading dynamics [21]. These approaches illustrate the complementary role of physical isolation techniques and advanced signal processing in ensuring scalable and reliable backscatter links.

*C. Key Contribution*

Although notable advances have been made, important challenges remain unresolved. Realizing reliable, multimedia-oriented backscatter systems requires integrating algorithmic innovation with practical hardware, while simultaneously achieving ultra-low power consumption and high data rates. This work addresses these gaps through the following key contributions:

- Reliable decoding requires rigorous treatment of residual phase-slope across pilot symbols was demonstrated. An analytically tractable but novel linearization approach is introduced to identify and cancel the phase-slope component in CE. It removes a key source of systematic bias while incurring no extra computational burden, rendering it highly suitable for ultra-low-power indoor backscatter applications.

- Whereas existing studies typically emphasize either algorithmic design or hardware implementation in isolation, this work delivers an integrated framework that spans theoretical modeling, signal processing, and prototyping. It provides design guidelines that close the gap between abstract estimation methods and real-world backscatter hardware, serving as a practical reference for subsequent research.

- The proposed architecture achieves simultaneous ultra-low-power consumption, high data rates, and extended communication distance. Both simulation and experimental validation confirm these performance gains, underscoring the system's ability to scale beyond conventional backscatter constraints.

- The proposed system reliably transmits multimedia data over a monostatic backscatter link. This contribution demonstrates not only the feasibility but also the practicality of backscatter as a medium for next-generation IoT applications requiring high throughput and robustness.

The remainder of this paper is organized as follows. Overall system model is described in Section II. In Section III, the proposed CE strategy is introduced, followed by Section IV and Section V, which details the system design and implementation. The performance evaluation, including both simulations and experimental results, is then presented in Section VI. Finally, the paper is concluded in Section VII.

## II. SYSTEM OVERVIEW

To enable robust multimedia communication in backscattering wireless systems, a monostatic backscatter multimedia data communication (BMC) system is designed, which tightly integrates custom-designed hardware components with a rigorously structured signal processing pipeline. The full framework of the proposed system is illustrated in Fig. 1, which shows the interaction between the software-defined radio (SDR)-based reader and the semi-passive backscatter tag. The key architectural decision to adopt a monostatic configuration—where both transmitting and receiving antennas are placed close to each other—reduces hardware complexity but introduces critical challenges such as self-interference, limited communication range, and severely attenuated signal-to-noise ratio (SNR) [22]. To address these challenges, a tailored hardware design is developed by constructing a high-gain planar Yagi–Uda antenna with split-ring resonators (SRRs) between the transmit (Tx) and receive (Rx) antennas. In parallel, in software, the signal processing pipeline is partitioned into two complementary stages: real-time pre-processing on the SDR and offline post-processing on the host PC. The backscatter tag employs the single-pole double-throw (SPDT) switch toggling between two impedance loads $Z_{L1}$ and $Z_{L2}$, thereby modulating the incident carrier with either on-off keying (OOK) or binary phase shift keying (BPSK). The tag's modulation is entirely passive and driven by a low-power control circuit, ensuring energy efficiency [23], [24]. The SDR transceiver transmits a continuous wave (CW) carrier signal and simultaneously captures the backscattered signal via the same antenna.

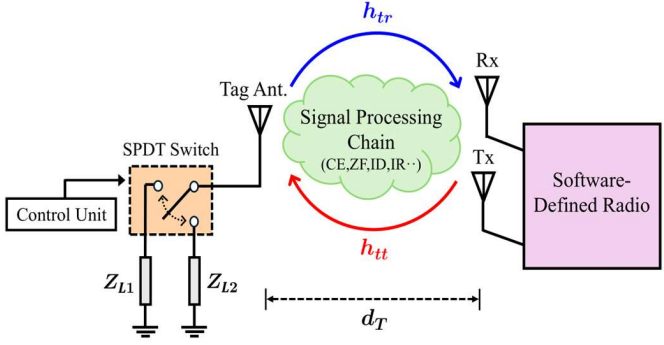

Fig. 1. Block diagram of the proposed monostatic BMC system ($d_T$: the distance between the tag and the transceiver).

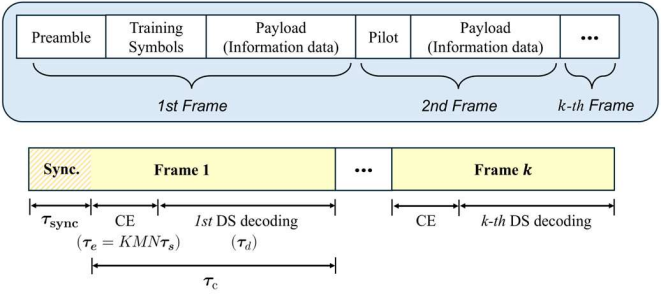

Fig. 2. Frame structure for channel estimation and information decoding.

Upon reception, the SDR executes the first stage of signal conditioning (pre-processing). The first operation is DC offset removal, which is essential to eliminate the baseband bias caused by the inherent characteristics of backscattering and internal hardware mismatches, especially prominent in monostatic architecture due to strong self-interference. Next, carrier synchronization and phase alignment are applied to counteract random phase shifts arising from round-trip propagation delay and the unknown phase state between the backscatter tag and reader. A frequency-domain low-pass filter (LPF) based on fast Fourier transform (FFT) is then used to isolate the narrowband backscattered signal from wideband noise and adjacent spectral leakage, yielding a cleaner signal representation for further processing [25].

In the post-processing stage, additional signal refinement and high-level decoding are performed. A second layer of DC offset removal further suppresses any residual baseline distortion. Pilot-aided CE is then conducted using both Least Squares (LS) and Linear Minimum Mean Square Error (LMMSE) techniques.

The proposed LS estimator provides a simple, phase-aware estimate of the backscatter channel coefficient using known pilot symbols, while the LMMSE estimator, which has a slightly more computational complexity than LS method, leverages second-order statistics to minimize estimation errors in noisy environments [26]. The inclusion of both estimators allows for comparative analysis of their estimation accuracy and subsequent impact on bit error rate (BER). Following process of CE, a Zero-Forcing (ZF) equalizer is applied the inverse of the estimated channel to mitigate the peak inter-symbol interference (ISI) distortion caused by the cascaded forward and backscattered channels $h_{tt}$ and $h_{tr}$.

After equalization, the signal is demodulated using either OOK or BPSK, followed by information decoding (ID) and image reconstruction (IR).

## III. System Model

### A. Backscatter Channel Model in SISO systems

In monostatic single-input single-output (SISO) backscatter communication systems, the received baseband signal is affected by two independent wireless channels: the forward channel from the reader to the tag, denoted $h_{tt}$, and the backscatter channel from the tag back to the reader, denoted $h_{tr}$. As the backscatter process is based on the passive reflection, the effective channel experienced at the receiver becomes the multiplicative product $h_{eq} \triangleq h_{tt}h_{tr}$. To accurately model this composite fading behavior, double-Rician fading channel model is adopted for both forward and backscattered channels. Specifically, $h_{tt} \sim \mathcal{CN}(V_{tt}, \sigma_{tt}^2)$ and $h_{tr} \sim \mathcal{CN}(V_{tr}, \sigma_{tr}^2)$, where $V_{tt}$ and $V_{tr}$ are the deterministic amplitudes of the line-of-sight (LoS) components, and $\sigma_{tt}^2$, $\sigma_{tr}^2$ are the diffuse powers. The corresponding Rician $K$–factors are given by $K_{tt} = |V_{tt}|^2/\sigma_{tt}^2$, $K_{tr} = |V_{tr}|^2/\sigma_{tr}^2$, which quantify the ratio of specular-to-diffuse power in each link [27]. The envelope of the equivalent channel $r \triangleq |h_{eq}| = |h_{tt}h_{tr}|$ follows a non-trivial probability distribution derived from the product of two independent Rician variables. Following [28, Eq. 6.66], the probability density function (PDF) of $r$ is expressed as:

$$p_h(r) = \frac{r}{\sigma_{tt}^2\sigma_{tr}^2}\exp(-(K_{tt}+K_{tr}))\sum_{i=0}^{\infty}\sum_{l=0}^{\infty}\frac{1}{(i!\,l!)^2}\left(\frac{rK_{tt}}{2\sigma_{tt}^2}\right)^i$$
$$\times \left(\frac{rK_{tr}}{2\sigma_{tr}^2}\right)^l\left(\frac{\sigma_{tt}}{\sigma_{tr}}\right)^{i-l}\mathbf{K}_{i-l}\left(\frac{r}{\sigma_{tt}\sigma_{tr}}\right), \qquad r \geq 0, \qquad (1)$$

where $\mathbf{K}_\nu(\cdot)$ is the modified Bessel function of the second kind with order $\nu$. The received instantaneous SNR is defined as:

$$\gamma = \frac{r^2E_s}{N_0}, \qquad (2)$$

by the change of variable $r = \sqrt{\gamma N_0/E_s}$, then the induced SNR probability is expressed as:

$$f_\gamma(\gamma) = \frac{1}{2}p_h\left(\sqrt{\frac{\gamma N_0}{E_s}}\right)\sqrt{\frac{N_0}{E_s\gamma}}, \qquad \gamma > 0, \qquad (3)$$

where $E_s$ denotes the energy of the symbol. In a multiplicative backscatter channel, moderate fading on either hop creates deep fades on $h_{eq}$, so $\gamma$ fluctuates much more strongly than in single-hop Rayleigh/Rician fading channels [29]. Consequently, the performance must be measured in expectation over the fading distribution rather than by instantaneous AWGN channels. The average-effective SNR is then obtained from the second moments of the constituent Rician fading:

$$\overline{\gamma}_{eff} = \mathbb{E}[|h_{eq}|^2]\frac{E_s}{N_0} = (\sigma_{tt}^2 + |V_{tt}|^2)(\sigma_{tr}^2 + |V_{tr}|^2)\frac{E_s}{N_0}$$

$$= (1 + K_{tt})(1 + K_{tr})\sigma_{tt}^2\sigma_{tr}^2\frac{E_s}{N_0}. \qquad (4)$$

Given a conditional error probability $P_e(\gamma)$, the average-effective BER is defined as:

$$\overline{P}_e = \int_0^\infty P_e(\gamma)f_\gamma(\gamma)\,d\gamma = \int_0^\infty P_e\left(\frac{E_s}{N_0}r^2\right)p_h(r)\,dr. \qquad (5)$$

In Section VI, the generic average-effective BER formulation in (5) is specialized by substituting the relevant conditional error laws for each detector. This process produces the average BER expressions for OOK and coherent BPSK, which form the basis of the receiver evaluation.

### B. Optimal Resource Allocation and Frame Design

Backscatter frames are designed to balance two competing requirements: (i) longer training reduces CE error and improves the effective SNR of the cascaded links, whereas (ii) shorter training increases the payload fraction and thereby the achievable data rate. The tension is acute under cascaded Rician fading because small estimation errors propagate strongly through the product channel and the reader estimates CSI only once per frame. Frame design is posed as an optimization problem that allocates time between pilots and data to satisfy a reliability target with minimum overhead. As shown in Fig. 2, let $\tau_{sync}$ denote the preamble duration, $\tau_e$ the training or tracking time, and $\tau_c$ the per-frame transmission time measured in symbol period units. In the proposed format a frame is composed of $K$ pilot slots, each slot spanning $M$ successive symbol periods per slot, and the pilot length is controlled by integer design parameter $N$. With symbol period $\tau_s$, the total training time is defined as $\tau_e \triangleq KMN\tau_s$, and the payload time within a frame is $\tau_d = \tau_c - \tau_e$ denotes the payload duration within a frame. Throughout this subsection, all times are expressed in seconds, so average energies scale linearly with the time durations under constant transmit power [30], [31].

Given transmitted pilot power $p_0$, the total pilot energy is defined as $s_0 = p_0\tau_e$. Without committing to a particular estimator, the single-frame Cramér-Rao lower bound (CRLB) is used to obtain a bound on the mean square error (MSE) of the CE,

$$\sigma_e^2(\tau_e) \geq \frac{N_0}{p_0\tau_e}, \qquad (6)$$

then map this error to the backscattered average-effective SNR by a standard one frame inversion model,

$$\gamma_{eff}^{CE}(\tau_e) \approx \frac{\overline{\gamma}_{eff}}{1 + \overline{\gamma}_{eff}\,\sigma_e^2(\tau_e)/\sigma_h^2}, \qquad (7)$$

where, $\overline{\gamma}_{eff}$ is in the absence of CE error given in (4) and $\sigma_h^2 = \mathbb{E}[|h_{eq}|^2]$. Substituting (6) into (7) yields a simple monotone improvement law:

$$\gamma_{eff}^{CE}(\tau_e) \approx \overline{\gamma}_{eff}\frac{\tau_e}{\tau_e + \alpha}, \qquad \alpha \triangleq \frac{\overline{\gamma}_{eff}N_0}{p_0\sigma_h^2}. \qquad (8)$$

This expression reveals a fundamental trade-off in frame design: increasing $\tau_e$ improves accuracy of CE, thereby enhancing the SNR $\gamma_{eff}^{CE}(\tau_e)$ (hence lowers $\overline{P}_e$); however, it also reduces the fraction of symbols available for payload transmission. The resulting effective spectral efficiency per frame is expressed as:

$$\mathcal{R}(\tau_e) = (1 - \rho)\log_2\left(1 + \gamma_{eff}^{CE}(\tau_e)\right)$$

$$= \left(1 - \frac{\tau_e}{\tau_c}\right)\log_2\left(1 + \gamma_{eff}^{CE}(\tau_e)\right). \qquad (9)$$

Denoting the pilot fraction by $\rho = \tau_e/\tau_c$. Hence, longer training increases reliability but decreases throughput, while shorter training improves throughput but degrades reliability. By explicitly quantifying this trade-off between reliability and throughput, the optimal pilot length can be determined by formulating an optimization problem [32]. Specifically, given a target reliability $\overline{P}_{tar}$ and a minimum spectral efficiency $\mathcal{R}_{min}$, $\tau_e$ is chosen to maximize $\mathcal{R}(\tau_e)$ subject to the constraints imposed by the target BER, total frame duration, and minimum rate:

$$\text{(P1)}: \max_{\tau_e \geq 0} \mathcal{R}(\tau_e)$$

$$\text{s.t. } \overline{P}_e\left(\gamma_{eff}^{CE}(\tau_e)\right) \leq \overline{P}_{tar},$$

$$\tau_{sync} + \tau_e \leq \tau_c,$$

$$\left(1 - \frac{\tau_e}{\tau_c}\right)R \geq R_{min}. \qquad (10)$$

The objective $\mathcal{R}(\tau_e)$ can be analyzed in closed form using (8)-(9). Let $g(\tau_e) = \gamma_{eff}^{CE}(\tau_e)$, then $\mathcal{R}(\tau_e) = (1 - \rho)\log_2(1 + g(\tau_e))$. One verifies that $g'(\tau_e) = \overline{\gamma}_{eff}\frac{\alpha}{(\tau_e + \alpha)^2} > 0$ and $g''(\tau_e) < 0$. The first-order condition of (P1) on the continuous relaxation is obtained by differentiation:

$$\frac{d\mathcal{R}}{d\tau_e} = -\frac{1}{\tau_c}\log_2(1 + g(\tau_e)) + \frac{1 - \tau_e/\tau_c}{\ln 2}\cdot\frac{g'(\tau_e)}{1 + g(\tau_e)}. \qquad (11)$$

The first term increases with $\tau_e$ whereas the second term decreases since $g' \propto (\tau_e + \alpha)^{-2}$, which proves unimodality and guarantees a unique stationary point $\hat{\tau}_e \in (0, \tau_c)$ that can be found by a one-dimensional search. Feasibility clips this stationary point to the interval induced by the constraints in (10).

Writing the reliability target as an SNR threshold $\Gamma_{tar} = \overline{P}_e^{-1}(\overline{P}_{tar})$, the bounds are expressed as:

$$\tau_{min} = \max\left\{1, \frac{\alpha\Gamma_{tar}}{\overline{\gamma}_{eff} - \Gamma_{tar}}\right\}, \qquad (12)$$

and the upper bound is

$$\tau_{max} = \min\left\{\tau_c - \tau_{sync}, \tau_c\left(1 - \frac{R_{min}}{R}\right)\right\}. \qquad (13)$$

Therefore, the optimal training time can be obtained as:

$$\tau_e^\star = \min\{\max\{\hat{\tau}_e, \tau_{min}\}, \tau_{max}\}. \qquad (14)$$

In implementation, $\tau_e^\star$ is converted to an integer pilot length by selecting a pilot placement and quantizing to symbols. Let the placement parameter be $\beta \in \{1, K\}$, where $\beta = 1$ denotes a single contiguous pilot burst and $\beta = K$ denotes per-slot pilots, with one block of length $N$ (in symbols) ahead of each of the $K$ slots. Let $\hat{\tau}_e$ be the unique stationary point of $\mathcal{R}(\tau_e)$, the training time realized by $N$ pilot symbols is $\hat{\tau}_e = \beta N \tau_s$. The implementable pilot length is therefore $N^\star = \lceil \tau_e^\star / \beta \tau_s \rceil$, which guarantees that the BER target used to obtain $\tau_e^\star$ is not violated. In cases where the frame budget is tight, $N^\star$ can be set to:

$$N^\star \leftarrow \max\{n \in \mathbb{Z}_+ : \tau_{sync} + \beta n \tau_s \leq \tau_c\}. \qquad (15)$$

Based on (14)-(15), the realized pilot fraction and payload duration are $\rho^\star = \hat{\tau}_e / \tau_c$ and $\tau_d^\star = \tau_c - \tau_{sync} - \hat{\tau}_e$.

## IV. SOFTWARE FRAMEWORK

After front-end conditioning in the SDR, the backscatter recording is stored as complex baseband samples at a fixed sampling rate. The post-processing stage on the host PC then executes a deterministic chain that removes residual baseline terms, estimates the composite backscatter channel consistent with the cascaded double-Rician fading, inverts that channel with a single frame equalizer, and finally decodes the information stream. The design goal is to deliver unbiased inputs to the estimators, conserve SNR at low operating levels, and keep each block's assumptions explicit and verifiable.

### A. DC Offset Removal

A CW reader transmits and receives through closely spaced antennas, which creates strong self-interference into the receiver. Local oscillator (LO) feedthrough and residual I/Q imbalance add a constant baseband term [33]. Quasi-static environmental reflectors contribute additional constant power independent of the tag switching. This DC component appears as a constant bias term in the discrete time complex baseband signal, and if not removed, it can significantly degrade subsequent CE and detection stages.

Let $r[n]$ denote the complex baseband signal obtained from the SDR after pre-processing. Over the operating interval the samples received are well described by:

$$y[n] = h_{eq}x[n] + d_{pre} + w[n]. \qquad (16)$$

Here, $x[n]$ is the digital baseband sequence induced by the tag modulation after carrier recovery, $d_{pre}$ is the dataset-wide DC term and $w[n]$ is zero mean AWGN. The offset evolves on a time scale much longer than a frame, therefore a single estimate for the entire recording is adopted. Received pilot samples are gathered in the index set $\mathcal{P}_k \subset \{n_0, \cdots, n_{k-1}\}$. The pilot sequence is designed to have zero symbol mean at the tag, $\mathbb{E}[x[n] \mid n \in \mathcal{P}_k] = 0$, and the model of the received samples implies $\mathbb{E}[y[n] \mid n \in \mathcal{P}_k] = d_{pre}$. Consequently, the DC offset is calculated by the mean of the pilot sample:

$$\hat{d}_{pos} \approx \frac{1}{T_k^{tot}} \sum_{n \in \mathcal{P}_k} y[n], \qquad (17)$$

where $T_k^{tot}$ denotes the total number of received samples. The estimator is unbiased and efficient; its mean and variance are given by $\mathbb{E}[\hat{d}_{pos}] = d_{pre}$, and $\sigma(\hat{d}_{pos}) = \sigma_w^2 / T_k^{tot}$. The debiased record is then expressed as:

$$\tilde{y}[n] = y[n] - \hat{d}_{pos}, \qquad (18)$$

which yields zero empirical mean after de-biasing and preserves the noise covariance assumed in the subsequent CE process.

### B. Optimal Residual Phase-Aware Channel Estimation

Reliable decoding in the monostatic SISO backscatter link requires an accurate estimate of the cascaded channel, as even small estimation errors propagate multiplicatively through the two-hop link. When the composite gain is unknown, the payload samples experience uncontrolled rotation and scaling, which degrades the post-equalization SNR and produces an error floor that persists even as payload energy increases. This challenge is especially severe in backscatter systems, where channel variations, such as residual phase drifts, accumulate across both hops.

To handle these issues, optimized LS or LMMSE estimators are developed, which extend the conventional approaches by incorporating a linearized model for intra-frame phase slope. Unlike these ordinary estimators that assume the channel remains strictly constant over the pilot block, the proposed phase-slope-robust versions explicitly model residual carrier drift or oscillator mismatch as a small linear phase perturbation. These robust estimators retain linearity, admit closed-form solutions, and eliminate the systematic bias that their ordinary counterparts suffer in the presence of phase drift.

After the DC removal across the entire received signal, received pilot samples within a coherence frame followed the flat SISO channel can be modeled as:

$$\tilde{y}_p[n] = h_{eq}e^{j\phi_1 n}x_p[n] + w[n], \qquad (19)$$

where $\phi_1$ is a residual phase-slope parameter, introduced by imperfections such as hardware mismatch, slight frequency offsets or oscillator instability. The exponential term $e^{j\phi_1 n}$ models a small linear phase drift across the pilot symbols. $x_p \triangleq (A_s - \xi_i)x_0$ represents the known pilot symbols, where $A_s$ is the antenna structural mode independent with a load and $\xi_i$ represents selected load reflection coefficients, and $x_0$ is the underlying pilot waveform. Notice that $A_s$ and $\xi_i$ are complex values, and the equivalent channel $h_{eq}$ is a single complex scalar that is constant over the pilot interval.

*1) LS Estimation Method:* Assuming that the residual phase-slope is small over a short pilot interval, then it can be approximated to $e^{j\phi_1 n} \approx 1 + j\phi_1 n$, which follows from the first-order Taylor series expansion. Providing the unknown lifted phase $\theta_0 \triangleq h_{eq}$, $\theta_1 \triangleq j\phi_1 h_{eq}$, the received samples reduce to a linear form,

$$\tilde{y}_p[n] \approx \theta_0 x_p[n] + \theta_1 n x_p[n] + w[n]. \tag{20}$$

Stacking the pilot samples into the observation vector $\mathbf{y_p} = \left[\tilde{y}_p[n]\right]_{n \in \mathcal{P}_k} \in \mathbb{C}^{k \times 1}$, and defining $\mathbf{x_p} = \left[x_p[n]\right]_{n \in \mathcal{P}_k} \in \mathbb{C}^{k \times 1}$, the design matrix is written as:

$$\mathbf{X_p} \triangleq \left[\mathbf{x_p} \ \mathbf{n} \odot \mathbf{x_p}\right], \tag{21}$$

with $\mathbf{X_p} \in \mathbb{C}^{k \times 2}$, $\mathbf{n} = [n]_{n \in \mathcal{P}_k}$, and $\odot$ denoting the elementwise product. The linear model with $\boldsymbol{\theta} = [\theta_0 \ \theta_1]^T$ can be obtained:

$$\mathbf{y_p} = \mathbf{X_p}\boldsymbol{\theta} + \mathbf{w}. \tag{22}$$

The robust phase-slope LS problem therefore formulates the optimization problem as below:

$$\widehat{\boldsymbol{\theta}}_{LS} = \arg\min_{\boldsymbol{\theta}} \left\| \mathbf{y_p} - \mathbf{X_p}\boldsymbol{\theta} \right\|_2^2, \tag{23}$$

whose first-order stationarity condition, the solution is obtained by the normal equations in closed-form,

$$\widehat{\boldsymbol{\theta}}_{LS} = \left(\mathbf{X_p^H X_p}\right)^{-1} \mathbf{X_p^H y_p}, \tag{24}$$

where $\mathbf{X_p^H}$ is Hermitian transpose of $\mathbf{X_p}$. To evaluate this efficiently and to expose the trade-offs designed pilots, five sufficient frame-level statistics are introduced: $s_0 \triangleq \sum_{n \in \mathcal{P}_k} |x_p[n]|^2$, $s_1 \triangleq \sum_{n \in \mathcal{P}_k} n |x_p[n]|^2$, $s_2 \triangleq \sum_{n \in \mathcal{P}_k} n^2 |x_p[n]|^2$, $t_0 \triangleq \sum_{n \in \mathcal{P}_k} x_p^*[n] \tilde{y}_p[n]$, $t_1 \triangleq \sum_{n \in \mathcal{P}_k} n x_p^*[n] \tilde{y}_p[n]$, where $s_0$ is the pilot energy, $s_1$ quantifies the time centering of that energy, $s_2$ quantifies its temporal spread, $t_0$ is the matched-filter between the pilot and the received samples, and $t_1$ is the same correlation by the sample index to capture the small linear phase across the pilot burst. With these definitions, the Gram matrix and right-hand side reduce to

$$\mathbf{X_p^H X_p} = \begin{bmatrix} s_0 & s_1 \\ s_1 & s_2 \end{bmatrix}, \qquad \mathbf{X_p^H y_p} = \begin{bmatrix} t_0 \\ t_1 \end{bmatrix}, \tag{25}$$

which are the only quantities required to obtain the closed-form estimator. The estimator is identifiable when the Gram matrix is positive definite, which in this notation is the single condition

$$\Delta \triangleq s_0 s_2 - s_1^2 > 0. \tag{26}$$

Under this condition, the solution of the normal equations is obtained by inverting the $2 \times 2$ matrix:

$$\begin{bmatrix} \widehat{\theta}_0 \\ \widehat{\theta}_1 \end{bmatrix} = \frac{1}{\Delta} \begin{bmatrix} s_2 & -s_1 \\ -s_1 & s_0 \end{bmatrix} \begin{bmatrix} t_0 \\ t_1 \end{bmatrix}. \tag{27}$$

Consequently, the LS estimated equivalent channel from the first component can be obtained:

$$\widehat{h}_{LS} \triangleq \widehat{\theta}_0 = \frac{s_2 t_0 - s_1 t_1}{\Delta}. \tag{28}$$

From (27), the small residual phase-slope component can be isolated from the normal-equation solution. A diagnostic estimate is therefore defined by normalizing the first-order coefficient with respect to the DC term and taking the imaginary part,

$$\widehat{\phi}_1 = \Im\left(\frac{\widehat{\theta}_1}{\widehat{\theta}_0}\right) = \Im\left(\frac{s_0 t_1 - s_1 t_0}{s_2 t_0 - s_1 t_1}\right), \tag{29}$$

where $\Im(\cdot)$ represents the imaginary part of the ratio, isolating the contribution associated with phase variation across the frame. This component merits separate analysis, as even a small residual slope disrupts phase stationarity, degrades coherent combining, and can inflate the phase contribution to estimation error. If this phase is ignored and the conventional LS estimator $\tilde{h}_{OLS} = t_0/s_0$ is used, a first-order expansion of $e^{j\phi_1 n}$ shows

$$\tilde{h}_{OLS} \approx h_{eq} + j\phi_1 h_{eq} \frac{s_1}{s_0} + \frac{1}{s_0} \sum_{n \in \mathcal{P}_k} x_p^*[n] w[n], \tag{30}$$

so the estimate is systematically biased by the term $j\phi_1 h_{eq} s_1/s_0$ whenever the pilot indices are non-centered in time. That bias would be carried into the ZF equalizer as a combined amplitude and phase error and would not disappear even at high nominal SNR. The proposed two-parameter LS formulation removes this issue because the second column $\mathbf{n} \odot \mathbf{x_p}$ explicitly models the small linear phase across the pilots, and the cross terms in the closed-form solution above cancel the effect of $s_1$ for any pilot placement that satisfies $\Delta > 0$.

The estimation accuracy follows directly from standard LS covariance analysis. With zero-mean noise of variance $\sigma_w^2$,

$$\text{Cov}(\widehat{\boldsymbol{\theta}}_{LS}) = \sigma_w^2 (\mathbf{X_p^H X_p})^{-1} = \frac{\sigma_w^2}{\Delta} \begin{bmatrix} s_2 & -s_1 \\ -s_1 & s_0 \end{bmatrix}, \tag{31}$$

and the variance of the LS estimation is written as:

$$\sigma_{e,LS}^2 = \frac{\sigma_w^2 s_2}{\Delta}. \tag{32}$$

These expressions expose the trade-off with frame design. For a fixed pilot energy $s_0$, centering the pilot indices so that $s_1 = 0$ maximizes $\Delta$ and minimizes $\text{Var}(\widehat{h}_{LS})$. Increasing the temporal span of the pilots increases $s_2$ and improves the conditioning through $\Delta = s_0 s_2 - s_1^2$, provided $s_1$ remains close to zero. These relations guide the choice of pilot locations when allocating training time inside a frame.

*2) Linear Minimum Mean Square Error Estimation Method:*
When second-order channel statistics are available, the estimator is obtained by minimizing the MSE over affine functions of the observations vector $\mathbf{y_p}$:

$$(\mathbf{M}, \mathbf{c})^\star = \widehat{\boldsymbol{\theta}}_{LM} = \arg \min_{\mathbf{M} \in \mathbb{C}, \mathbf{c} \in \mathbb{C}} \mathbb{E}\left[ \left| \boldsymbol{\theta} - \left( \mathbf{M}^H \mathbf{y_p} + \mathbf{c} \right) \right|^2 \right], \quad (33)$$

where the expectation is over the channel, the residual phase term, and the receiver noise. A Gaussian prior is adopted to summarize slow statistics across frames: $\boldsymbol{\theta} \sim \mathcal{CN}(\boldsymbol{\mu_\theta}, \boldsymbol{\Lambda_\theta})$, where $\boldsymbol{\mu_\theta} = [\mu_h \ \mu_1]^T$ and $\boldsymbol{\Lambda_\theta} = \operatorname{diag}(\sigma_h^2, \sigma_1^2)$, with $\mu_1 \approx 0$ since assuming that the residual slope fluctuates around zero. Under this linear-Gaussian model, the solution of (33) is the posterior mean,

$$\widehat{\boldsymbol{\theta}}_{LM} = \boldsymbol{\mu_\theta} + \boldsymbol{\Lambda_\theta} \mathbf{X_p}^H \left( \sigma_w^2 \mathbf{I}_k + \mathbf{X_p} \boldsymbol{\Lambda_\theta} \mathbf{X_p}^H \right)^{-1}$$
$$\times \left( \mathbf{y_p} - \mathbf{X_p} \boldsymbol{\mu_\theta} \right). \quad (34)$$

Starting from the lifted model (22) with (21), the normal equations depend only on the Gram matrix $\mathbf{X_p}^H \mathbf{X_p}$, already obtained in (25), and on the data-prior correlation vector $\mathbf{X_p}^H (\mathbf{y_p} - \mathbf{X_p} \boldsymbol{\mu_\theta})$. Using the $s_0, s_1, s_2, t_0$, and $t_1$, which have already been defined for the LS, the expressions can be written as:

$$\mathbf{X_p}^H (\mathbf{y_p} - \mathbf{X_p} \boldsymbol{\mu_\theta}) = \begin{bmatrix} t_0 - \mu_h s_0 - \mu_1 s_1 \\ t_1 - \mu_h s_1 - \mu_1 s_2 \end{bmatrix}, \quad (35)$$

which separates the information from the received matched filters $(t_0, t_1)$ and the subtraction of what would be expected from the prior $\boldsymbol{\mu_\theta}$. To keep the algebra compact and to make the role of pilot energy, pilot spread, prior, and noise explicit, introduce the scalars: $b_{00} \triangleq \sigma_h^{-2} + \sigma_w^{-2} s_0$, $b_{11} \triangleq \sigma_1^{-2} + \sigma_w^{-2} s_2$, $b_{01} \triangleq \sigma_w^{-2} s_1$, $a_0 \triangleq \sigma_w^{-2} t_0 + \sigma_h^{-2} \mu_h$, $a_1 \triangleq \sigma_w^{-2} t_1 + \sigma_1^{-2} \mu_1$, and the positive determinant is simplified as:

$$\Delta_B \triangleq b_{00} b_{11} - b_{01}^2 > 0. \quad (36)$$

The coefficients $b_{00}$, $b_{11}$, $b_{01}$ collect, respectively, prior precision and noise-weighted pilot moments into a two-by-two normal matrix, while $a_0, a_1$ collect the prior–data correlations. With these definitions the posterior mean of the LMMSE problem reduces to a single two by two inversion,

$$\widehat{\boldsymbol{\theta}}_{LM} = \left( \boldsymbol{\Lambda_\theta}^{-1} + \sigma_w^{-2} \mathbf{X_p}^H \mathbf{X_p} \right)^{-1} \left( \sigma_w^{-2} \mathbf{X_p}^H \mathbf{y_p} + \boldsymbol{\Lambda_\theta}^{-1} \boldsymbol{\mu_\theta} \right)$$
$$= \frac{1}{\Delta_B} \begin{bmatrix} b_{11} & -b_{01} \\ -b_{01} & b_{00} \end{bmatrix} \begin{bmatrix} a_0 \\ a_1 \end{bmatrix}. \quad (37)$$

Consequently, the LMMSE estimated equivalent channel gain can get from the first component,

$$\widehat{h}_{LM} = \left( \widehat{\boldsymbol{\theta}}_{LM} \right)_1 = \frac{b_{11} a_0 - b_{01} a_1}{\Delta_B}. \quad (38)$$

The posterior covariance follows from the same normal matrix and equals,

$$\operatorname{Cov}\left( \widehat{\boldsymbol{\theta}}_{LM} \right) = \left( \boldsymbol{\Lambda_\theta}^{-1} + \sigma_w^{-2} \mathbf{X_p}^H \mathbf{X_p} \right)^{-1}$$
$$= \frac{1}{\Delta_B} \begin{bmatrix} b_{11} & -b_{01} \\ -b_{01} & b_{00} \end{bmatrix}, \quad (39)$$

and the variance of the LMMSE estimation is expressed as:

$$\sigma_{e,LM}^2 = \frac{b_{11}}{\Delta_B} \quad (40)$$

These formulations show exactly which design parameters matter. The denominators $\Delta_B$ and $b_{11}$ increase with pilot energy $s_0$ and with the temporal spread $s_2$ of the pilots, both of which improve conditioning, while the cross term $s_1$ only appears inside $b_{01}$ and cancels out through $\Delta_B$. Thus, the lifted LMMSE remains insensitive to a non-centered pilot placement that would bias a scalar LMMSE [34].

For comparison, the conventional scalar LMMSE estimator that ignores the second column of $\mathbf{X_p}$ would solve the upliftied problem $\mathbf{y_p} = h_{eq} \mathbf{x_p} + \mathbf{w}$ and give the familiar shrinkage,

$$\widetilde{h}_{OLM} = \frac{\sigma_h^2 s_0}{\sigma_h^2 s_0 + \sigma_w^2 s_0} \frac{t_0}{s_0} + \frac{\sigma_w^2}{\sigma_h^2 s_0 + \sigma_w^2} \mu_h. \quad (41)$$

If a small linear phase persists across the pilots and the index set of the pilots is not centered ($s_1 \neq 0$), a first-order expansion of the samples shows that $\widetilde{h}_{OLM}$ inherits the deterministic bias term $j \phi_1 h_{eq} s_1 / s_0$. The lifted LMMSE developed removes that bias by estimating the residual phase coefficient jointly with the channel, while still providing Bayesian shrinkage toward the calibrated prior when the pilot length is short or the SNR is low.

Both estimators are designed using a simple basis expansion, thereby avoiding significant computational overhead while attenuating the residual phase component.

### C. Zero-Forcing Equalization

In each frame, the optimal designated pilot interval of duration $\tau_e$ provides an estimated scalar channel gain $\widehat{h}$ for the forward and backscattered links. The subsequent payload samples of that same frame are then equalized by a single complex gain, defined as the ZF equalizer $p \triangleq \widehat{h}^{-1}$, and applied symbol-by-symbol to form the normalized channel sequence before the decision over payload indices only. The goal is to suppress the dominant single frame ISI attributable to the product channel. When the information data satisfies the flat SISO model in symbol units, it can be expressed as $y_d = h_{eq} x_d + w_d$, then the equalized symbols are defined as $z_d \triangleq p y_d$. Using the data model, the equalized stream decomposes as:

$$z_d = x_d + \left( \frac{h_{eq}}{\widehat{h}} - 1 \right) x_d + \frac{w_d}{\widehat{h}}. \quad (42)$$

A desired term identical to the transmitted symbols, a self-interference term that is proportional to the mismatch $h_{eq} - \widehat{h}$, and a noise component amplified by the factor $1/|\widehat{h}|$. This decomposition is specific to the monostatic backscatter link because the end-to-end distortion inside a frame is a single phasor; consequently, the ZF is the unique linear operation that removes that phasor without altering timing or bandwidth [35]. Performance ties directly to the resource allocation variables in (6)–(14). If the post-equalization SNR expression of (8) is re-evaluated by inserting the estimation error power $\sigma_e^2(\tau_e)$ produced by the chosen estimator and $\tau_e$, the resulting average-effective SNR reduces to the same law used to design the frame.

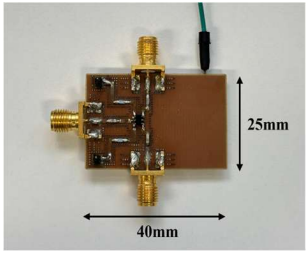

Fig. 3.    Designed semi-passive tag for load modulation.

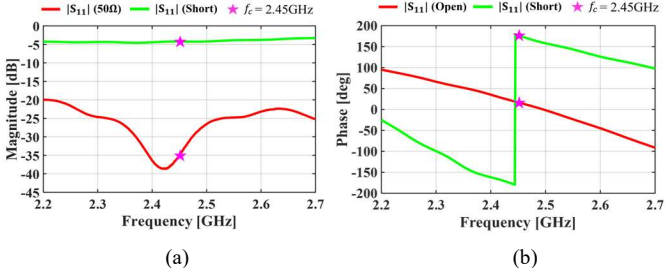

Fig. 4.    Measured $S_{11}$ of the proposed load modulator (semi-passive tag) (a) magnitude and (b) phase.

## V. HARDWARE IMPLEMENTATION

### A. Tag Design and Modulation Schemes

The backscatter tag plays a critical role in encoding information onto an incident carrier wave through a low-power load modulation technique. The proposed tag employs a single antenna interfaced with a controllable SPDT switch to alternate between two distinct complex loads, $Z_{L1}$ and $Z_{L2}$, enabling binary modulation. The reflection behavior of the tag is governed by the impedance mismatch between the antenna impedance $Z_a$ and the selected load $Z_L$, resulting in a reflection coefficient $\xi_i$ defined by:

$$\xi_i = \frac{Z_{Li} - Z_a^*}{Z_{Li} + Z_a}, \qquad i = 1, 2, \tag{43}$$

where $Z_a^*$ denotes the complex conjugate of the antenna impedance. The modulation scheme is implemented by toggling between two distinct impedance states, each corresponding to a unique load reflection coefficient. For OOK modulation, one of the impedances is matched to the antenna impedance so that $\xi_i \approx 0$. The other impedance is deliberately mismatched to induce strong reflection with $|\xi_i| \approx 1$. This amplitude-based scheme is straightforward to implement and well suited to energy-constrained tags. For BPSK modulation, the two impedance states are chosen so that their reflection coefficients satisfy $\xi_{i1} = -\xi_{i2}$. This yields constant envelope backscatter with a phase separation of $\pi$.

To realize this modulation scheme, an ultra-low-power SPDT switch is designed and implemented to enable efficient impedance control for binary modulation in the backscatter tag. The switch employs the HMC221BE device from Analog Devices, which offers a typical insertion loss of 0.7 dB, contributing to the overall low loss performance of the circuit. The fabricated circuit, consuming only 5 μW, is constructed on a compact 40 mm × 25 mm PCB using a 0.2 mm thick FR-4 substrate, which has a relative dielectric constant $(\varepsilon_r)$ of 4.14 and loss tangent (tan $\delta$) of 0.02 as shown in Fig. 3. The reflection characteristics of the proposed switch are measured using a calibrated vector network analyzer (VNA), and the results are shown in Fig. 4. The measured reflection magnitude, $|S_{11}|$ exhibits clear discrimination between the two load states as shown in Fig. 4(a). When terminated with a standard 50 Ω load representing the reflective mode, the $|S_{11}|$ reaches approximately –35.0 dB at the center frequency of 2.45 GHz, indicating strong reflection. In contrast, the short-circuit termination yields a magnitude of –4.2 dB, corresponding to increased absorption. These results confirm the intended impedance contrast between the two states, necessary for reliable amplitude-based modulation. On the other hand, Fig. 4(b) shows the respective phase response under open and shorted load conditions. At 2.45 GHz, the measured reflection phase is approximately 16.8° in the shorted state and 177.5° in the open state, yielding a differential phase of 160.7°.

Accordingly, further analysis is required to identify the non-idealities leading to the deviation from the ideal 180° phase shift and complete absorption ($|S_{11}| \approx 0$). Specifically, the –4.2 dB reflection in the absorptive state is attributed to the insertion loss of the SPDT switch (~1.1 dB), losses in SMA connectors and cables (~1.5 dB), inherent insertion loss in the short termination itself, and parasitic capacitance arising from the switch layout and measurement setup. In the operating frequency, the non-negligible reactance of capacitive elements further reduces the reflection coefficient due to partial energy absorption. As for the phase deviation, the short-to-open reflection phase difference fails to reach a perfect 180° because parasitic capacitances and inductances within the switch circuit introduce signal distortion. Even with a measured ~161° differential, robust antipodal modulation is maintained under practical backscatter conditions. These measurements confirm that the proposed switching circuit delivers performance close to ideal for OOK and BPSK modulation in the backscatter tag.

### B. Tx/Rx High Isolation Antenna Design

The designed antenna is shown in Fig. 5 and 6. The antenna is primarily designed to strengthen the useful backscattered signal while suppressing self-interference in the compact reader. In the proposed platform, the SDR transmits CW signals and simultaneously receives the tag's modulated response on closely spaced ports. The short antenna separation induces strong electrical coupling between the transmit and receive feeds, allowing carrier leakage into the receive path. This leakage reduces the effective SNR by forcing the front-end to operate with limited gain, generating a strong deterministic component after down-conversion, and distorting the statistics used for DC-offset removal and pilot-aided CE. Consequently, a high-isolation antenna system is essential to ensure reliable operation of the entire signal processing chain. The detailed design parameters of the proposed antenna are listed in Table I.

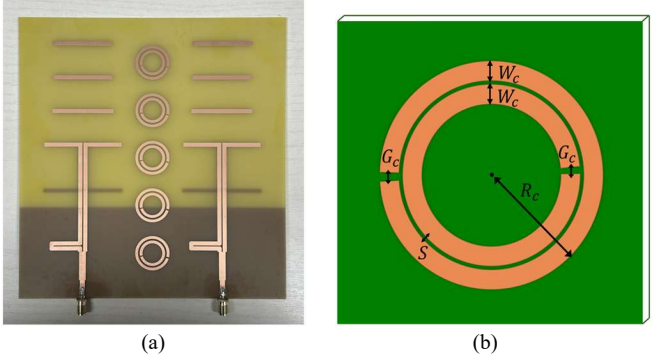

Fig. 5.    (a) Fabricated 2×1 planar Yagi-Uda antenna array and (b) geometry of the SRR for isolation improvement.

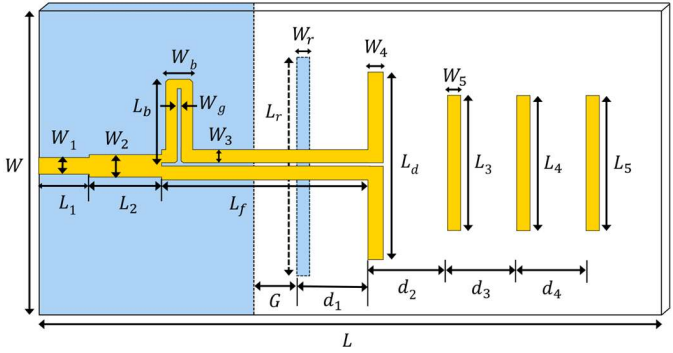

Fig. 6.    Configuration of the proposed antenna.

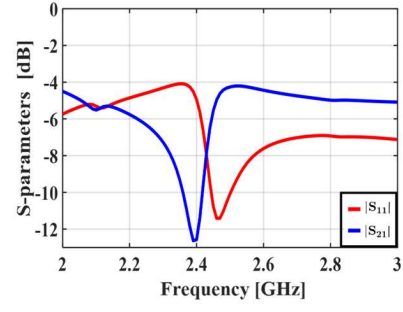

Fig. 7.    Frequency response of the designed SRR: $|S_{11}|$ & $|S_{21}|$.

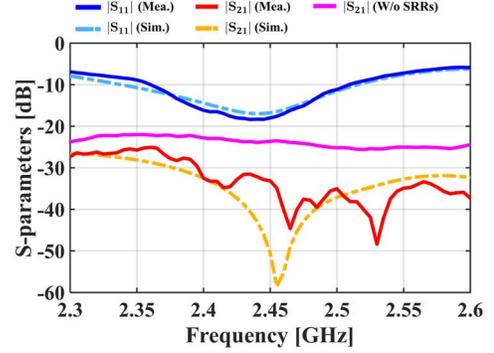

Fig. 8.    Simulated and measured S-parameters of the proposed antenna: $|S_{11}|$ & $|S_{21}|$.

TABLE I
DESIGN PARAMETERS OF ANTENNA & SRR (UNIT: MM)

| $W$ | 76 | $W_b$ | 5.5 | $L_3$ | 37 | $L_r$ | 48 |
|-----|-----|-------|-----|-------|-----|-------|-----|
| $W_1$ | 3 | $W_g$ | 0.5 | $L_4$ | 36 | $d_1$ | 28 |
| $W_2$ | 4 | $W_r$ | 2.5 | $L_5$ | 36 | $d_2$ | 20.5 |
| $W_3$ | 2.5 | $L$ | 170 | $L_b$ | 17 | $d_3$ | 20.5 |
| $W_4$ | 2.5 | $L_1$ | 11 | $L_d$ | 46.5 | $d_4$ | 20.5 |
| $W_5$ | 2.5 | $L_2$ | 17 | $L_f$ | 63.75 | $G$ | 8.75 |
| $G_c$ | 1 | $R_c$ | 11.5 | $W_c$ | 1.5 | $S$ | 0.5 |

The SRR array is integrated between the transmit and receive regions of the antenna board to suppress the dominant near-field electric coupling. A SRR stores electric energy in the gaps and magnetic energy in the loops. At resonance, the structure exhibits a high surface impedance to tangential electric fields, effectively suppressing lateral field lines that could otherwise couple energy from the transmit feed into the receive feed. The magnitude of $|S_{21}|$ has a deep transmission notch, and $|S_{11}|$ exhibits the companion resonance around 2.4~2.5 GHz as shown in Fig. 7. The resonators are first tuned at the unit cell level, then realized as an array on FR-4 substrate and placed between the two apertures. The measured S-parameters in Fig.

8 indicate that incorporating the resonators reduces the inter-port coupling $|S_{21}|$ to below −30 dB, which is approximately 10 dB better than the case without resonators. At the same time, the input match $|S_{11}|$ remains better than −10 dB, confirming that the isolation enhancement does not come at the expense of matching or bandwidth.

Nevertheless, isolation by itself does not guarantee reliable operation. To counteract the substantial double-hop path loss and low backscattered signal power, the reader and the tag should use high-gain antennas directed along their LoS to maximize link reliability. To achieve this, planar Yagi–Uda antennas are adopted for both the tag and the reader. The antenna's geometry concentrates illumination onto the tag, enhances the front-to- back ratio, and selectively receives energy from the desired direction, thereby directly improving the SNR and increasing the communication distance. As shown in Fig. 9, the measured radiation patterns demonstrate a realized gain of 7 dBi at 2.45 GHz, with good agreement to simulations in both principal planes, thereby validating the array's end-fire radiation. The experimental performance characterization was carried out in an anechoic chamber, as depicted in Fig. 10.

### C. Software-Defined Radio as the reader: Pre-Processing

The reader employs a SDR platform in which the field programmable gate array (FPGA) provides unified control of the transmit and receive signal paths, as illustrated in Fig. 11. During data transmission, the digital controller assembles the frame sequence and drives the digital up-converter and digital-analog converter (DAC), which in turn feed the RF chain [36], [37].

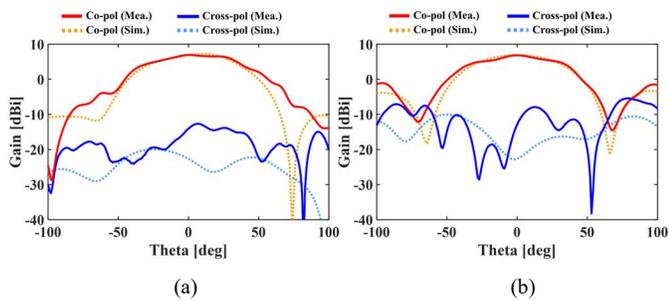

Fig. 9. Simulated and measured gain radiation patterns at 2.45 GHz for (a) $\phi = 0°$ in $xz$-plane and (b) $\phi = 90°$ in $yz$-plane.

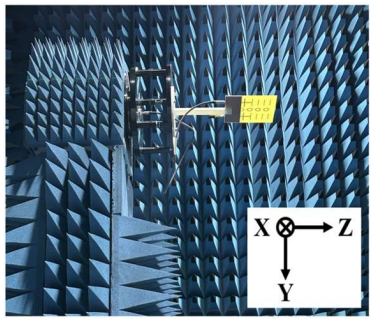

Fig. 10. The antenna measurement setup in the anechoic chamber.

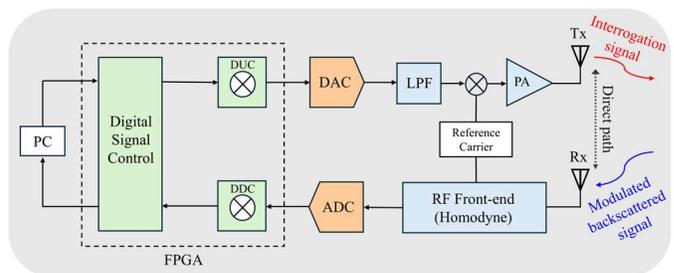

Fig. 11. Block diagram of the SDR reader implementing pre-processing through digital signal control on FPGA.

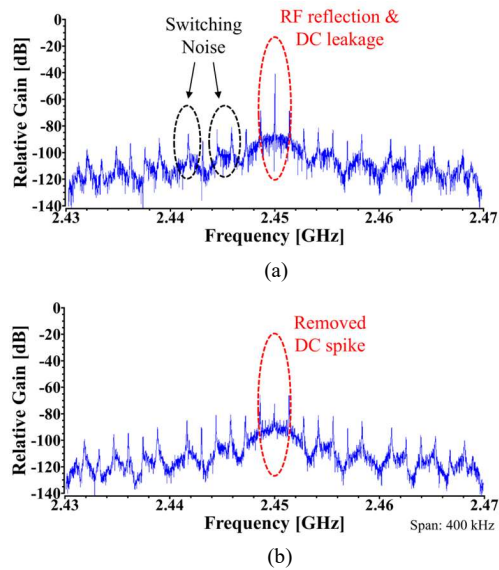

Fig. 12. Frequency spectrum of the receiver during data reception at the center frequency of 2.45 GHz: (a) before and (b) after DC offset removal.

The first operation removes the large direct-current component that is inherent to a monostatic backscatter reader. In a homodyne receiver, the LO frequency is identical to the RF carrier frequency due to direct conversion. Any leakage of the LO signal or the RF carrier into the mixer's input path results in a self-mixing component, producing a deterministic DC term at the baseband signal. At the same time, the low-noise amplifier (LNA) in front of the mixer can further amplify this term. The result is a large DC spike in the spectrum and a constant bias in the time domain. The FPGA tracks and removes this term on the entire recording so that this one-shot correction eliminates the carrier line around the center frequency and adjusts the position of the constellation. During data reception, the spectrum prior to correction is shown in Fig. 12(a), where the carrier leakage and other spurious peaks are visible near 2.45 GHz. After the initial processing stage of the received data, the DC spike is suppressed as presented in Fig. 12(b). The remaining narrow spectral harmonics originate from the tag switch. As the data rate grows, steeper current and voltage transitions in the switch generate stronger harmonic content and induce non-linear mixing, which manifest as small peaks symmetrically located around the carrier frequency.

After the DC offset suppression, the complex baseband coming out of the ADC still exhibits a rapid constellation rotation. In the proposed system, this rotation is caused by a small frequency mismatch of the LO carrier between the transmitter that illuminates carrier to the tag and the receiver. Although the Tx and Rx are disciplined to the same reference clock, their phase-locked loop (PLL) frequency synthesizers operate independently. This mismatch introduces a carrier frequency offset (CFO) $\Delta f$ that imprints a linearly growing phase term across every frame and destroys the phase reference required by coherent detection. Residual CFO imposes a continuous phase rotation as illustrated in Fig. 13(a). To compensate for this effect, the raw discrete-time complex baseband signal is modeled as follows:

$$y_0[n] = (h_{eq}x_0[n] + d_0)e^{j(2\pi\Delta f nT_s + \Phi_0)} + w_0[n], \quad (44)$$

where $x_0[n]$ is the backscattered samples, $d_0$ is deterministic DC component, $w_0[n]$ is circular complex Gaussian noise, $T_s$ is sampling period, and $\Phi_0$ is an arbitrary initial phase. After sampling, the residue appears as a common complex rotation that multiplies both the desired backscatter component and the direct-path leakage.

The receiver computes the first-lag autocorrelation of the raw samples, $u[n] = y_0[n]y_0^*[n-1]$. The average statistic $\bar{u}[n]$ is then obtained over a short window of $W$ consecutive samples starting at index $n_0$, which isolates the CFO-induced phase increment. Under the standard assumption that adjacent data samples are nearly uncorrelated at lag one, a condition valid for OOK and BPSK with modest excess bandwidth, the cross terms vanish yielding $\mathbb{E}\{u[n]\} = p_0e^{j2\pi\Delta fT_s}$. Hence, the phase of $\bar{u}[n]$ directly reveals the CFO phase advance $2\pi\Delta fT_s$, independent of the unknown data symbols and the constant leakage amplitude. The CFO is estimated as below:

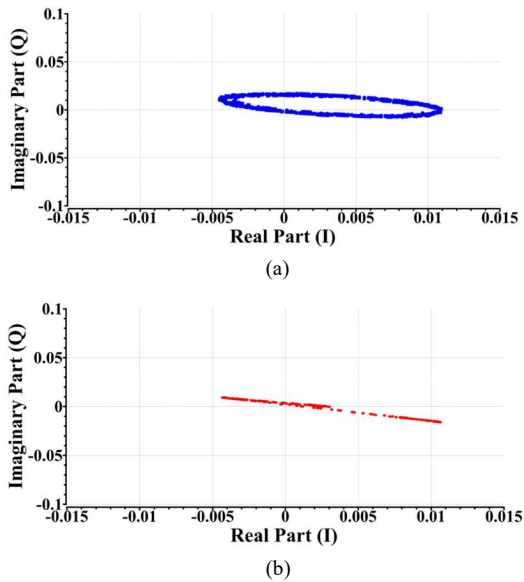

Fig. 13. Constellation diagram of the received data: (a) before and (b) after CFO compensation for enabling coherent detection.

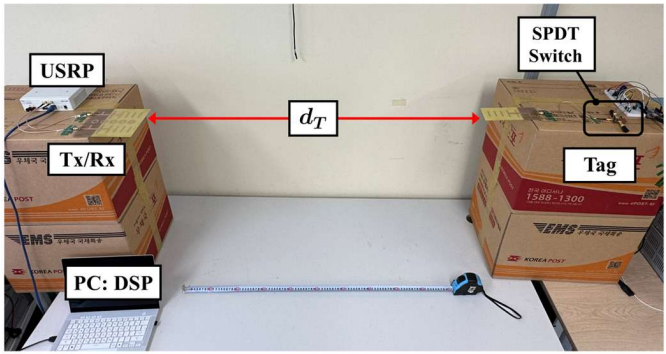

Fig. 14. Indoor experimental setup of the proposed BMC system: $d_T = 1$ m.

$$\widehat{\Delta f} = \frac{1}{2\pi T_s} \arg(\bar{u}[n]), \quad (45)$$

where $\arg(\cdot)$ denotes the phase angle of a complex variable in radians. With the estimated $\widehat{\Delta f}$, the received samples are multiplied by a complex sinusoidal wave of opposite rotation. The corrected signal is thus expressed by:

$$y[n] = y_0[n] e^{\{-j(2\pi\widehat{\Delta f}(n-n_0)T_s) + \widehat{\Phi}_0\}}, \quad (46)$$

where $\widehat{\Phi}_0 = \arg(y_0[n_0])$ serves as a reference for the absolute phase within the frame. This derotation procedure achieves carrier synchronization by transforming the constellation from a rotating ring into a phase-aligned line, as shown in Fig. 13(b). After compensation, the data stream conforms to the pre-processing model in (16). Consequently, CFO correction stabilizes the constellation, allowing subsequent CE and ZF equalization to operate on a phase-stationary record, which in turn enables reliable decoding. The improvement is clearly visible in the observed constellation diagram of Fig. 13(b).

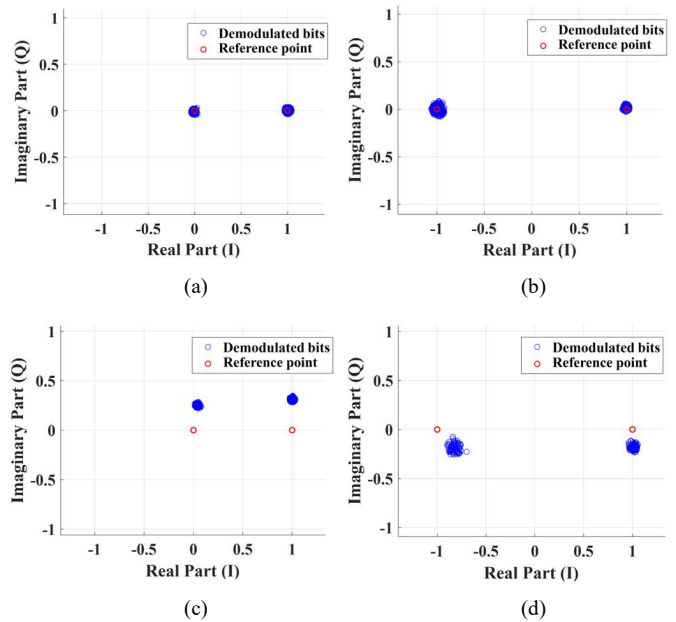

Fig. 15. Constellation diagram after the signal processing chain for OOK and BPSK with and without an integrated SRRs: (a) OOK w/ SRRs, (b) BPSK w/ SRRs, (c) OOK w/o SRRs, and (d) BPSK w/o SRRs.

## VI. EXPERIMENTAL RESULTS

In this section, the performance evaluation of the entire signal processing chain in monostatic backscatter links in an indoor experimental setup (Fig. 14) is discussed. First, the error vector magnitude (EVM) after the full signal processing chain is evaluated for OOK and BPSK. EVM quantifies the residual modulation error after equalization as the root-mean-square (RMS) distance between the equalized samples and the ideal constellation, normalized by the reference symbol power and reported in percent.

$$\text{EVM}_{\text{RMS}}[\%] = 100 \sqrt{\frac{\frac{1}{N}\sum_{k=1}^{N}|y_{ks} - s_{ks}|^2}{\frac{1}{N}\sum_{k=1}^{N}|s_{ks}|^2}}. \quad (47)$$

Here, $s_{ks}$ denotes the reference symbol and $y_{ks}$ the equalized sample at the decision instant after the full processing chain. With the SRRs integrated at the antenna, the measured EVM is 2.97% for OOK and 4.02% for BPSK. These results reflect higher isolation between the transmit and the receive path in the monostatic reader. The slightly larger EVM for BPSK is consistent with its phase sensitivity because residual phase error directly perturbs the decision boundary, whereas OOK is primarily amplitude-driven. However, this difference can be considered within the acceptable error margin. Without the SRRs, the EVM rises to 34% for OOK and 22% for BPSK. In this configuration, transmit leakage at the antenna dominates the receiver input, leaving a residual carrier at baseband even after digital suppression; the impact becomes more severe for higher-order modulation schemes where tighter amplitude and phase accuracy are required and the symbol-error probability increases significantly for a given EVM. Constellation diagram of the proposed BMC system with and without SRRs are shown in Fig. 15.

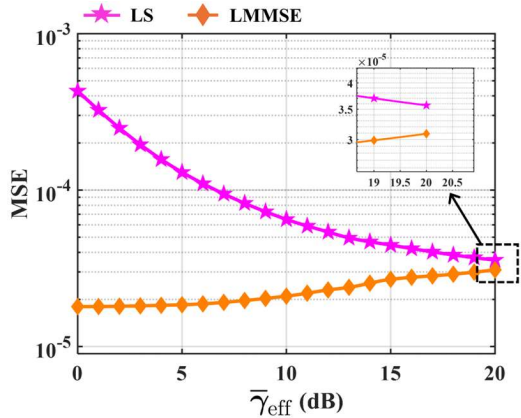

Fig. 16. MSE performance of LS and LMMSE estimation accuracy over varying SNR ($\overline{\gamma}_{\text{eff}}$).

The estimation accuracy of LS and LMMSE is evaluated in terms of MSE, as shown in Fig. 16. When $\overline{\gamma}_{\text{eff}} > 5$ dB, both estimators achieve MSE values below $10^{-4}$, demonstrating that reliable CSI can be obtained for backscatter links. In the low and moderate $\overline{\gamma}_{\text{eff}}$ regime, the LMMSE estimator in (38) outperforms the LS method in (28) by one to two orders of magnitude. This advantage arises from the shrinkage effect, where the LMMSE effectively combines the LS with prior channel statistics to suppress the variance of the pilot matched-filter statistic, thereby reducing estimation error for the same pilot energy. As $\overline{\gamma}_{\text{eff}}$ increases, the LS error continues to decrease because it is unbiased and dominated only by noise, while the LMMSE curve gradually flattens due to a small residual bias induced by shrinkage. Consequently, at high $\overline{\gamma}_{\text{eff}}$ or with sufficiently long pilot sequences, the LMMSE weight approaches unity and both estimators converge, making LS an equally effective yet computationally simpler alternative.

Thus, while the MSE provides a useful aggregate metric, the conventional form, $\mathbb{E}\left\{\left|\hat{h} - h_{eq}\right|^2\right\}$, in fact conflates two distinct contributions: (i) amplitude mismatch and (ii) phase ambiguity. When these sources are mixed, it can be unclear whether poor performance arises from magnitude error or residual phase rotation, and even a constant phase offset can artificially increase the reported MSE despite accurate amplitude estimation. To address this, the error can be decomposed into separate magnitude and phase components. Specifically, let $\hat{r} \triangleq |\hat{h}|$, and $\Delta\psi \triangleq \arg(\hat{h}) - \arg(h_{eq})$ so that $\Delta\psi$ represents the residual phase difference between the estimate and the true channel. By the law of cosines, the instantaneous squared error can be expressed as:

$$\left|\hat{h} - h_{eq}\right|^2 = (\hat{r} - r)^2 + 4\hat{r}r\sin^2\left(\frac{\Delta\psi}{2}\right). \quad (48)$$

Taking expectations and using linearity of the expectation operator, the following decomposition is obtained:

$$\mathbb{E}\left\{\left|\hat{h} - h_{eq}\right|^2\right\} = \underbrace{\mathbb{E}\left[(\hat{r} - r)^2\right]}_{\text{Magnitude term}} + \underbrace{\mathbb{E}\left[4\hat{r}r\sin^2\left(\frac{\Delta\psi}{2}\right)\right]}_{\text{Phase term}}. \quad (49)$$

Notice that the relationship holds identically for every realization of $(r, \hat{r}, \Delta\psi)$, and therefore also under expectation. The decomposition preserves overall comparability with the conventional MSE while providing diagnostic insight: for instance, improvements in synchronization or phase tracking manifest as a reduction in the phase component, whereas pilot energy, pilot length, and estimator design primarily influence the magnitude component [38].

To analyze BER performance, the BER probability expressions for OOK and BPSK from (5) is derived. The BER of OOK on the instantaneous SNR is defined as $P_e^{\text{OOK}}(\gamma) = Q\left(\sqrt{\frac{\gamma}{2}}\right)$. Substituting into (5) gives a single integral over the product-Rician envelope [39]. The average BER is

$$\overline{P}_b^{\text{OOK}} = \int_0^\infty Q\left(\sqrt{\frac{E_s}{2N_0}}r\right)p_h(r)\,dr. \quad (50)$$

However, due to the complexity of $p_h(r)$, the closed-form expression is not feasible, and the Meijer-G function $\mathbf{G}_{p,q}^{m,n}$ is used as an approximation expressed in terms of $\overline{\gamma}_{\text{eff}}$. Equation (50) can be simplified to:

$$\overline{P}_b^{\text{OOK}} \approx \frac{1}{12}\mathbf{G}_{1,3}^{3,0}\left(\frac{\overline{\gamma}_{\text{eff}}}{4}\,\middle|\,\begin{matrix}1\\1, \frac{1}{2}, 0\end{matrix}\right). \quad (51)$$

For coherent BPSK, $P_e^{\text{BPSK}}(\gamma) = Q(\sqrt{2\gamma})$, where $Q(x) = \frac{1}{\sqrt{2\pi}}\int_x^{+\infty}\exp\left(-\frac{t^2}{2}\right)dt$ denoting the Gaussian Q-function. Likewise, using (5) with the $p_h(r)$, closed-form is not available; a compact Gaussian Q-function approximation proposed by authors in [40] is employed, and the average BER of coherent BPSK can be written as follows:

$$\overline{P}_b^{\text{BPSK}} \approx \frac{1}{12}\mathbf{G}_{1,3}^{3,0}\left(\overline{\gamma}_{\text{eff}}\,\middle|\,\begin{matrix}0\\1, 1, 1\end{matrix}\right)$$
$$+ \frac{1}{4}\mathbf{G}_{1,3}^{3,0}\left(\frac{4}{3}\overline{\gamma}_{\text{eff}}\,\middle|\,\begin{matrix}0\\1, 1, 1\end{matrix}\right). \quad (52)$$

Equations (51) and (52) are derived in Appendix.

Monte Carlo simulation with $10^5$ results indicate that both modulations exhibit BER significantly above the AWGN benchmark due to the multiplicative nature of the channel. As a result, BER curves shift rightward relative to the theoretical limits. The $K$–factor of the Rician distribution, which denotes the power ratio between the specular and scattered components, is set to 0, 7, 10, and 14 dB, where $K = 0$ corresponds to Rayleigh fading. As $K$ increases, the performance improves by reinforcing the specular component, yet the gap to AWGN persists since both links must be simultaneously strong. For both modulation schemes, a reference baseline is included for comparison: maximum-likelihood (ML) detection in the presence of CSI as the value of $K$.

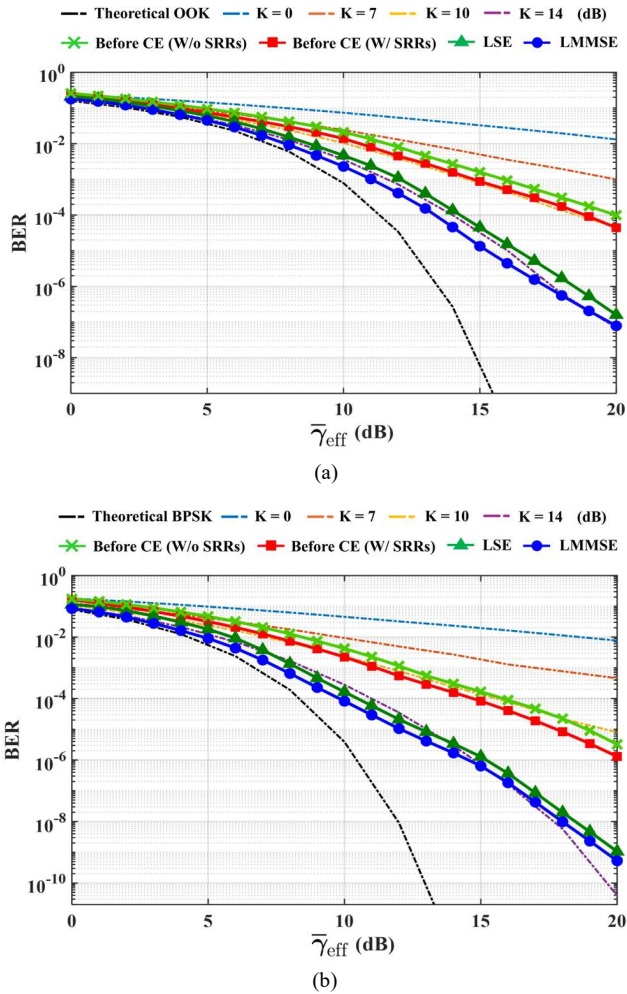

Fig. 17. BER performance versus $\overline{\gamma}_{\text{eff}}$ under the backscatter fading channel according to Rician K-factor: (a) OOK and (b) BPSK.

To quantify the benefit of CE, the achievable SNR during the ID phase was estimated. The backscattered SNR is determined by the accuracy of the available channel knowledge. In practice, CE methods are applied, but the resulting estimates always include errors. Consequently, the achievable SNR cannot be directly expressed in closed form and must be approximated using the error variance of the corresponding estimator.

Using the presence of CSI in (4), the backscattered SNR during ID from (8) can be defined. The parameter $\alpha$ is defined as $\alpha \triangleq \overline{\gamma}_{\text{eff}} N_0 / p_0 \sigma_h^2 = \overline{\gamma}_{\text{eff}} / \overline{\gamma}_e$, which represents the ratio between info SNR and pilot SNR with $\overline{\gamma}_e = p_0 \sigma_h^2 / N_0$. Substituting the LS estimation error variance (32) into (8), the achievable effective SNR during ID is expressed as:

$$\overline{\gamma}_{\text{eff}}^{\text{LS}} = \overline{\gamma}_{\text{eff}} \frac{\overline{\gamma}_e \Delta / s_2}{\overline{\gamma}_e \Delta / s_2 + \alpha}. \tag{53}$$

From the LMMSE, the posterior error variance of the channel gain is given by (40). Therefore, it becomes

$$\overline{\gamma}_{\text{eff}}^{\text{LM}} = \overline{\gamma}_{\text{eff}} \frac{\overline{\gamma}_e \Delta_{\text{B}} / b_{11}}{\overline{\gamma}_e \Delta_{\text{B}} / b_{11} + \alpha}. \tag{54}$$

In the case of OOK, the BER depends on the factor $\sqrt{\overline{\gamma}_{\text{eff}}^{\text{CE}} / 2}$, which directly ties system reliability to the achievable effective SNR. When channel estimation is absent, the error variance remains high and significantly suppresses the effective SNR. Once LS or LMMSE estimation is employed, this error contribution is markedly reduced, which effectively elevates $\overline{\gamma}_{\text{eff}}^{\text{CE}}$ and shifts the BER curve by roughly 4~5 dB, as illustrated in Fig. 17(a). A further enhancement arises from the inclusion of SRRs, which strengthen both the forward and backscattered paths. This reinforcement increases the overall channel power, thereby scaling both $\overline{\gamma}_{\text{eff}}$ and $\overline{\gamma}_e$, and accounts for an additional 2 dB gain. Moreover, because LMMSE leverages prior knowledge of channel statistics, it produces a smaller estimation error compared with LS, resulting in a further about 2 dB improvement. Collectively, these effects explain why the OOK system exhibits layered performance gains as CE and SRRs are introduced and as more advanced estimation methods are adopted.

For BPSK, the BER incorporates $\sqrt{2\overline{\gamma}_{\text{eff}}^{\text{CE}}}$, which amplifies the sensitivity to estimation quality. Here, the role of CE becomes even more decisive: suppressing estimation error through LS or LMMSE raises the achievable SNR enough to deliver $\approx$ 5 dB reduction in the required SNR at a fixed BER, as shown in Fig. 17(b). The gain from SRRs mirrors the OOK case in mechanism but not in magnitude of impact; by boosting the effective channel variance, the SRRs contribute a consistent 1~2 dB improvement across the BER curve. The distinction between LS and LMMSE is again visible: by minimizing the MSE with the aid of statistical priors, LMMSE estimation yields about 2 dB additional benefit relative to LS. Thus, while the sources of improvement are structurally similar to those in OOK, their effect is magnified in BPSK due to the coherent detection process, making the performance gap with and without CE more pronounced.

The visual reconstruction images as depicted in Fig. 18(a)~(f) further corroborate these analytical findings. Without CE, the restored images under OOK and BPSK exhibit severe distortion, which is consistent with the degraded effective SNR observed in the BER curves.

Table II provides a system-level comparison of reported backscatter communication systems. The monostatic ASK system in [42] achieves 67 kbps with energy consumption of 100 nJ/bit over a 5 m range, while the bistatic OOK/FSK system in [43] supports 1 kbps over 130 m but requires 20 μJ/bit, making it unsuitable for energy-constrained IoT nodes. Ambient backscatter approaches demonstrate varied trade-offs: the 4-PSK system in [44] achieves only 20 kbps over 0.76 m without pilot-based estimation, while the differential BPSK (DBPSK) design in [45] supports a higher throughput of 1 Mbps over 15 m but at a much higher energy cost of 1.8 nJ/bit. In contrast, the proposed platform demonstrates a balanced and scalable design, achieving 500 kbps at 1 m with an energy efficiency of 320 pJ/bit. The proposed design integrates both LS and LMMSE estimators, enabling robust operation under fading and residual phase drift. These results establish the practicality of the proposed platform as one of the first multimedia-capable backscatter IoT systems that jointly

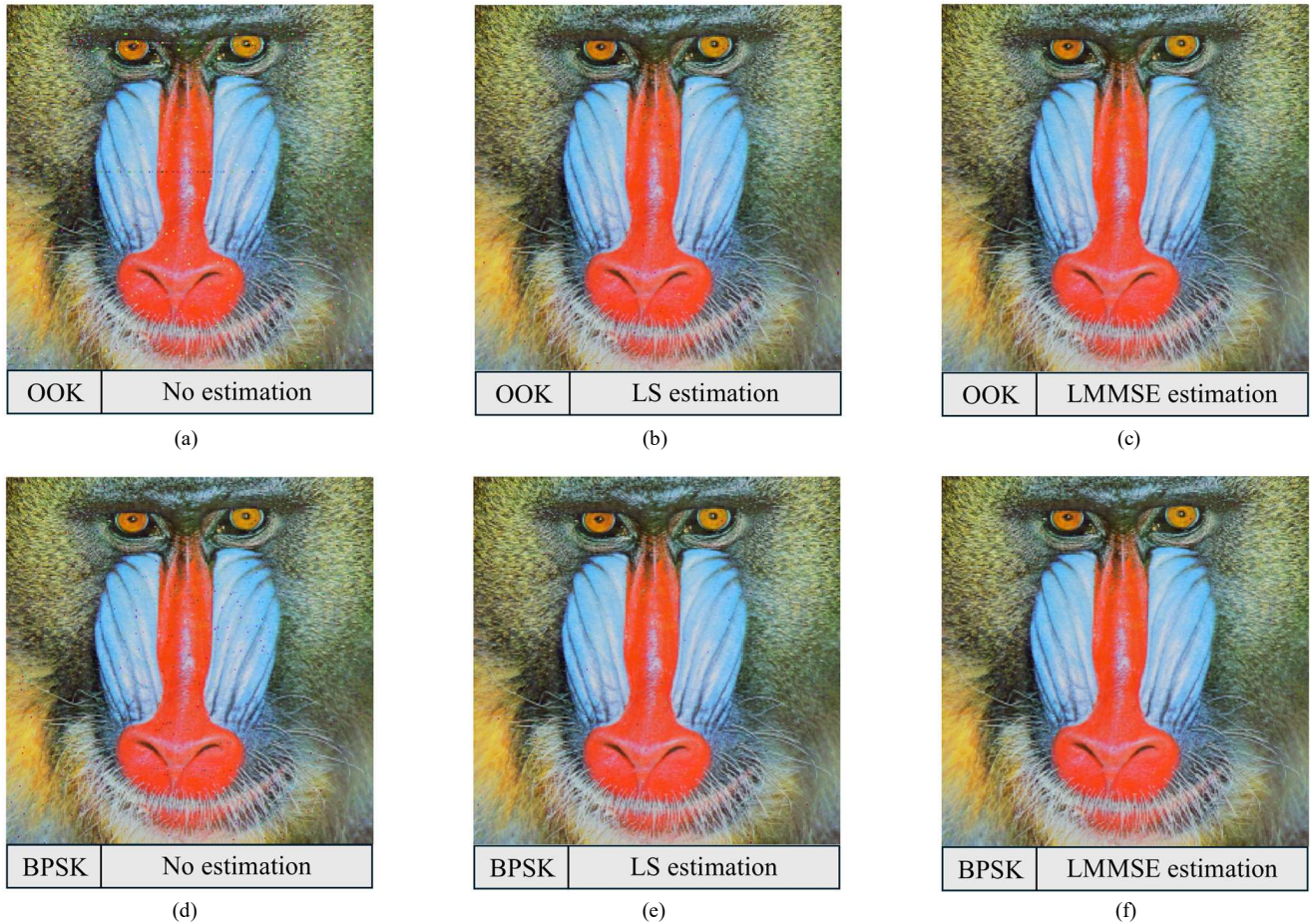

Fig. 18.    Over-the-air demonstration of the proposed BMC system under $\overline{\gamma}_{\text{eff}} = 15 \ dB$: (a) OOK w/o CE at BER $= 9.3 \times 10^{-4}$, (b) OOK w/ LS at BER $= 5.1 \times 10^{-5}$, (c) OOK w/ LMMSE at BER $< 10^{-9}$, (d) BPSK w/o CE at BER $= 9.1 \times 10^{-5}$, (e) BPSK w/ LS at BER $= 1.4 \times 10^{-6}$, and (f) BPSK w/ LMMSE at BER $< 10^{-9}$.

TABLE II
PERFORMANCE COMPARISON TABLE OF SYSTEM-LEVEL BACKSCATTER SYSTEMS

| Ref. | Config. | Modulation scheme | Energy efficiency | Max. Data rate | Max. Range | Channel estimation |
|---|---|---|---|---|---|---|
| [41] | Monostatic (MIMO) | BPSK | $\sim$ pJ/bit | N. A. | 13 cm | LS |
| [42] | Monostatic | ASK | 100 nJ/bit | 67 kbps | 5 m | ✗ |
| [43] | Bistatic | OOK, FSK | 20 µJ/bit | 1 kbps | 130 m | ✗ |
| [44] | Ambient | 4-PSK | N. A. | 20 kbps | 0.76 m | ✗ (Variance est.) |
| [45] | Ambient | DBPSK | 1.8 nJ/bit | 1 Mbps | 15 m | ✗ |
| This work | Monostatic | OOK, BPSK | 320 pJ/bit | 500 kbps | 1 m | LS, LMMSE |

achieves high throughput, ultra-low energy consumption, and estimation robustness.

## VII. CONCLUSION

This work demonstrates that reliable multimedia data communication is feasible in ultra-low-power backscatter settings, as required by modern IoT applications. A cost-effective and energy-efficient monostatic system is presented, integrating a custom tag, a single-antenna SDR reader, planar Yagi-Uda antennas, and a resonator network that enhances isolation without degrading radiation performance. The signal chain is rigorously derived and fully realized in practice: pre-processing at the reader stabilizes the carrier and suppresses deterministic offsets, while post-processing at the host applies optimized CE and ZF algorithms prior to ID and IR. The estimators are tailored to the double-Rician fading channel and mitigate residual phase-slope across each frame, thereby enabling robust OOK and BPSK operation even at low SNR. End-to-end experiments confirm that the theoretical model, hardware design, and software pipeline align to achieve consistent multimedia recovery. The same architecture naturally points to several extensions. Energy harvesting at the tag could eliminate batteries and support sustained multimedia links under practical power budgets. Multi-antenna readers and tags could further enhance performance through spatial selectivity, diversity, and interference rejection, achieving the reliability and throughput typically associated with MIMO while keeping the reader compact. These directions preserve the cost-effective and energy-efficient philosophy of the present design while expanding its applicability to broader IoT deployments.

## APPENDIX

### A. Deviation of Equation (51)

Let $r = |h_{eq}| \triangleq |h_{tt} h_{tr}|$ denote the product Rician envelope within a frame. For BER performance of OOK with optimal threshold, the conditional BER at instantaneous SNR $\gamma$ is

$$P_e^{\text{OOK}}(\gamma) = Q\left(\sqrt{\frac{\gamma}{2}}\right), \qquad (55)$$

where $\gamma = E_s/N_0\, r^2$. Averaging over the envelope yields

$$\bar{P}_b^{\text{OOK}} = \int_0^\infty Q\left(\sqrt{\frac{E_s}{2N_0}} r\right) p_h(r)\, dr. \qquad (56)$$

Let $\alpha \triangleq E_s/(2N_0)$, then a compact form that follows from Craig's identity

$$Q(\sqrt{\alpha} r) = \frac{1}{\pi} \int_0^{\pi/2} \exp\left(\frac{\alpha r^2}{2\sin^2\theta}\right) d\theta. \qquad (57)$$

Inserting (57) into (56) and exchanging the order of expectation and integration gives

$$\bar{P}_b^{\text{OOK}} = \frac{1}{\pi} \int_0^{\pi/2} \mathbb{E}[\exp(-s(\theta) r^2)] d\theta, \qquad (58)$$

with $s(\theta) = \alpha/2\sin^2\theta$, $\bar{P}_b^{\text{OOK}}$ is the $\theta$-average of the Laplace transform of $r^2$. Since, $\mathcal{L}_{r^2}(s) = 1 - s\mathbb{E}[r^2] + \mathcal{O}(s^2)$ as $s \to 0$, the small SNR expansion becomes

$$\bar{P}_b^{\text{OOK}} = \frac{1}{2} - \frac{\alpha}{4}\mathbb{E}[r^2] + \mathcal{O}(\alpha^2) = \frac{1}{2} - \frac{1}{4}\bar{\gamma}_{\text{eff}} + \mathcal{O}(\bar{\gamma}_{\text{eff}}^2). \quad (59)$$

At high SNR, the integral inherits the stretched-exponential tail $\sim z^{1/4} e^{-2\sqrt{z}}$ with $z = \bar{\gamma}_{\text{eff}}/4$, governed by the Bessel kernel. The Meijer-G kernel $\mathbf{G}_{1,3}^{3,0}\left(z \,\Big|\, 1, \frac{1}{2}, 0\right)$ has precisely the same first-order low-SNR term and the same high-SNR tail. Matching the low-SNR slope determines the scale factor, which gives the compact approximation

$$\bar{P}_b^{\text{OOK}} \approx \frac{1}{12} \mathbf{G}_{1,3}^{3,0}\left(\frac{\bar{\gamma}_{\text{eff}}}{4} \,\Big|\, 1, \frac{1}{2}, 0\right). \qquad (60)$$

This single-parameter expression preserves the dominant dependence on $\bar{\gamma}_{\text{eff}}$.

### B. Deviation of Equation (52)

For coherent BPSK the conditional error probability is defined as:

$$P_e^{\text{BPSK}}(\gamma) = Q(\sqrt{2\gamma}). \qquad (61)$$

Hence the average BER is simplified to:

$$\bar{P}_b^{\text{BPSK}} = \int_0^\infty Q(\sqrt{2\gamma}) f_\gamma(\gamma) d\gamma$$
$$= \int_0^\infty Q\left(\sqrt{\frac{2E_s}{N_0}} r\right) p_h(r)\, dr, \qquad (62)$$

where $p_h(r)$ is pdf of the double-Rician envelope.

Because the Q-function is evaluated at $x = \sqrt{2\gamma} \geq 0$, the Gaussian Q-function approximation by Chiani, Dardari, and Simon can be used as follows:

$$Q(x) \approx \frac{1}{12} e^{-\frac{x^2}{2}} + \frac{1}{4} e^{-\frac{2x^2}{3}}, \qquad x \geq 0. \qquad (63)$$

Their two-term exponential fit is uniformly tight for all non-negative arguments, with very small relative errors. As a result, it converts the average of $Q(\sqrt{2\gamma})$ over the double-Rician envelope into a weighted sum of Laplace transform of $r^2$:

$$\bar{P}_b^{\text{BPSK}} \approx \frac{1}{12} \mathbb{E}\left[e^{-\frac{E_s}{N_0} r^2}\right] + \frac{1}{4} \mathbb{E}\left[e^{-\frac{4E_s}{3N_0} r^2}\right]. \qquad (64)$$

The Laplace transform $\mathcal{L}_{r^2}(s) = \mathbb{E}[e^{-sr^2}]$ is well captured while matching both the low SNR slope $\mathcal{L}_{r^2}(s) = 1 - s\mathbb{E}[r^2] + \mathcal{O}(s^2)$ and the high SNR stretched-exponential tail induced by the Bessel by the Meijer-G function

$$\mathcal{L}_{r^2}(s) = \mathbf{G}_{1,3}^{3,0}\left(s\bar{\gamma}_{\text{eff}} \,\Big|\, \begin{matrix} 0 \\ 1, 1, 1 \end{matrix}\right), \qquad (65)$$

whose series at the origin is $1 - s\mathbb{E}[r^2] + \mathcal{O}(s^2)$ once the argument is normalized by $\bar{\gamma}_{\text{eff}}$.

Substituting (63) into (62) yields the desired closed-form, average-effective SNR expression:

$$\bar{P}_b^{\text{BPSK}} \approx \frac{1}{12} \mathbf{G}_{1,3}^{3,0}\left(\bar{\gamma}_{\text{eff}} \,\Big|\, \begin{matrix} 0 \\ 1, 1, 1 \end{matrix}\right)$$
$$+ \frac{1}{4} \mathbf{G}_{1,3}^{3,0}\left(\frac{4}{3}\bar{\gamma}_{\text{eff}} \,\Big|\, \begin{matrix} 0 \\ 1, 1, 1 \end{matrix}\right). \qquad (66)$$